# TYLOSIN ABATEMENT IN WATER BY PHOTOCATALYTIC PROCESS

Laoufi N.A[1,4], Alatrache A.[2,4], Pons M.N.[3], Zahraa O.[4]

[1] USTHB, Faculté de Génie Mécanique et de Génie des Procédés. Département de Génie Chimique et de Cryogénie. B.P. 32, 16111 El-Alia, Algérie
[2] Ecole Supérieure des Sciences et Techniques de Tunis, 5 Av. Taha Hussein, BP 56, Bab Menara, 1008, Tunisie
[3] Laboratoire des Sciences du Génie Chimique, CNRS, Nancy Université, ENSIC, 1 rue Grandville, BP 20451, F-54001 Nancy Cedex, France
[4] Département de Chimie-Physique des Réactions, UMR 7630 CNRS-INPL, Nancy Université, ENSIC, 1 rue Grandville, BP 20451, F-54001 Nancy Cedex, France

**Abstract**

The photocatalytic degradation of tylosin has been studied using immobilized titanium dioxide as catalyst. T he processes of degradation and reduction of tylosin was examined, and the activity is dominantly dependent on the surface coverage of the catalysts, The Langmuir-Hinshelwood model is satisfactorily obeyed at initial time and in the course of the reaction. These results suggest the feasibility of a photocatalytic system on the elimination of such antibiotic.

## 1 Introduction

The pollution of ground water and rivers by organic pollutants is of great concern and requires a constant improvement of water treatment. Some of these pollutants are used as pharmaceuticals and are present in hospital and domestic wastewater as well as in solid human and animal waste. Carried away to wastewater treatment plants through sewer networks, they can be released in the aquatic environment. Note that antibiotics administered to animals contribute to the fluxes of water contaminants. One of the best known of these antibiotics is tylosin, a powerful macrolide used in veterinary medicine.

Conventional water treatment includes adsorption and oxidation. Adsorption on activated carbon is efficient but requires a treatment of the adsorbent in order to eliminate the adsorbed pollutant. Oxidative treatment by chlorine or chlorine dioxide is not totally efficient and leads to the formation of toxic products. Advanced oxidation processes (AOPs), using hydrogen peroxide or ozone are more efficient but may be costly. An alternative cheaper AOP is photocatalysis, which has been tested on some antibiotics [1,2].

The photocatalyst is mainly a semiconductor, oxide or sulphide, and titanium dioxide $TiO_2$ is the most interesting from the point of view of efficiency and stability. Absorption of a photon in near UV range promotes an electron $e^-_{cb}$ in the conduction band, which produces a hole $h^+_{vb}$ in the valence band. If separated, these two species can migrate to the catalyst surface and act as reducer (oxygen reduction in superoxide ion) and oxidant (water oxidation in hydroxyl radical for example, or direct oxidation of the reactant) thus regenerating the catalyst electronic population. Highly reactive species such as hydroxyl radical can react on adsorbed organic molecules and abstract a hydrogen atom and therefore induce an oxidation.

## 2 Materials and methods

In the present work, the feasibility of tylosin degradation by photocatalysis has been investigated in a lab-scale photocatalytic reactor (Fig.1), where the polluted solution is recycled over a glass slide on which $TiO_2$ particles have been deposited and fixed as explained elsewhere [3]. The illumination is provided by a UV lamp emitting at 365 nm. Degradation kinetics was monitored by UV-visible spectroscopy, mass spectrometry (MS), HPLC-MS and dissolved organic matter quantification. The titanium dioxide (Degussa P25) was mainly anatase. According to the manufacturer's specifications [4], the elementary particle in dry powder form was approximately spherical in shape and the particle size was approximately 20 nm. The specific surface area, as measured from $N_2$ adsorption at 77 K, was 44 $m^2$ $g^{-1}$, in agreement with the manufacturer's specification. The size distribution of the particles suspended in water was measured by laser diffraction on a Malvern Mastersizer apparatus; the average suspended particle size was 5μm. Two other titanium dioxide samples were also used in this study, PC500 and PC 105 from Millennium. Tylosin tartrate was a commercial reagent purchased from Sigma-Aldrich and was used without further purification. The water used for all solutions and slurries was ion exchanged prior to use.

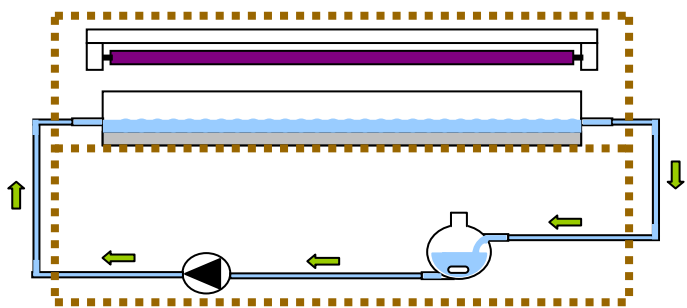

Figure 1: Recycling of contaminated water over the irradiated $TiO_2$ photocatalyst

## 3   Results

The three catalysts have shown an interesting activity as shown in Figures 2 and 3. In fact, analyses by mass spectrometry at different time intervals (data not shown) have confirmed the breakage of the tylosin molecule, under the conjugated effect of titanium dioxide and irradiation.

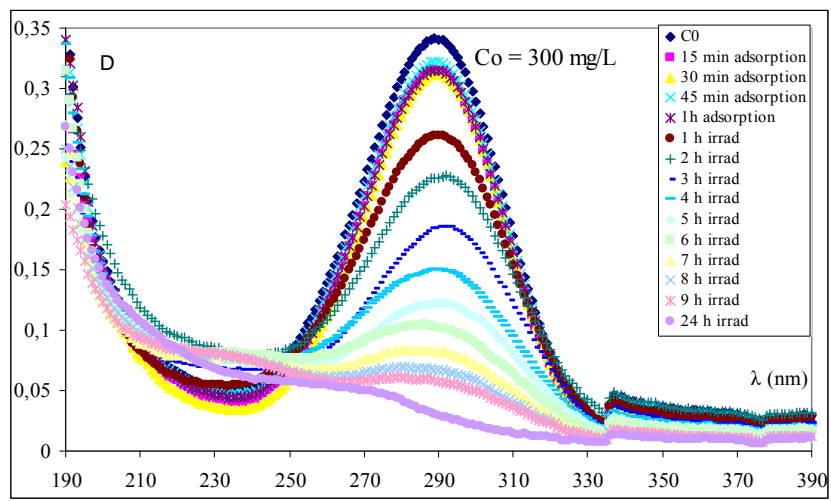

Figure 2.  Degradation of tylosin on Degussa P25.

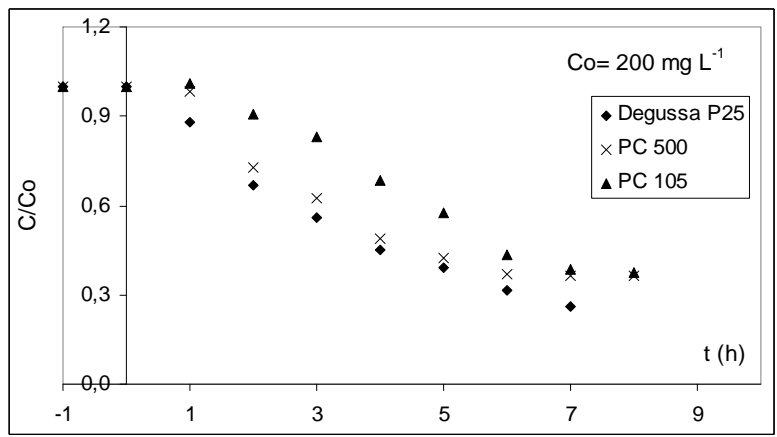

Figure 3.  Effect of titanium dioxide type

### 3.1 **Photocatalytic activity.**

As the best activity seems to be obtained with $TiO_2$ P25, it was of importance to focus the study on the influence of various parameters such as tylosin initial concentration on the degradation with this photocatalyst. Figure 4 shows that the kinetic of tylosin disappearance is enhanced at lower concentration. This phenomenon has been also observed in liquid [5] and gas phase [6], which could be explained by the surface hindrance of $TiO_2$ in presence of high concentration of tylosin.

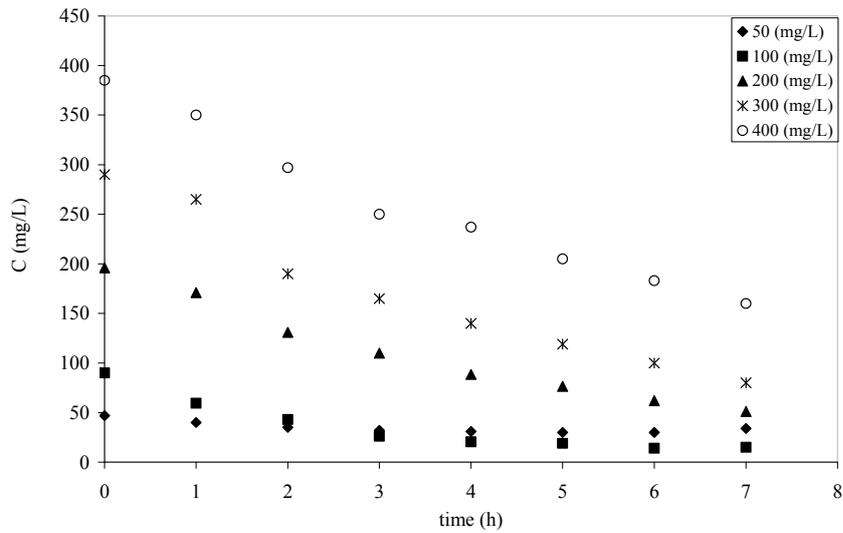

Figure 4, Kinetic of Tylosin degradation at various initial concentrations

### 3.2 **Order with respect to tylosin.**

The effect of the initial tylosin concentration C0 on the initial rate of photocatalytic degradation, r0, is shown in Figure 5. The rate is roughly proportional to the concentration, which corresponds to an initial kinetic order of one with respect to tylosin. This is in agreement with the simple Langmuir-Hinshelwood (LH) model.

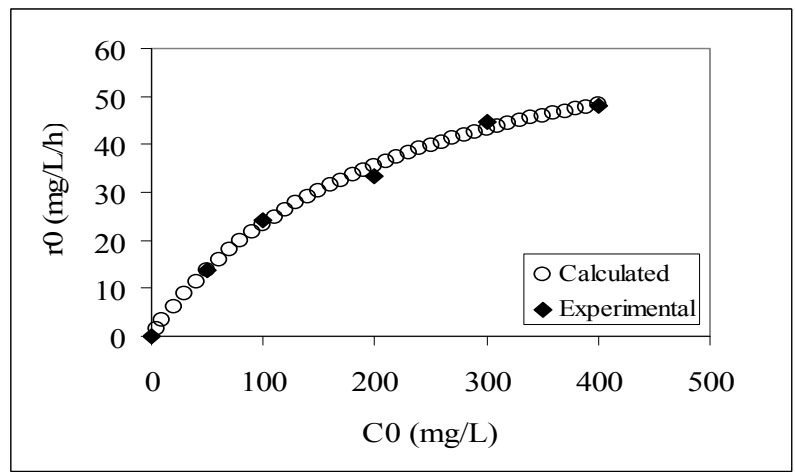

Figure 5 Initial rate variations with the initial tylosin concentration

In the LH model, the reaction of degradation follows a rate determining step where an adsorbed molecule react with a reactive transient such as an OH• radical or a hole. If this attack is a minor process of disappearance of the reactive transient, the rate of reaction is proportional to the surface coverage by the pollutant. Assuming a simple Langmuir model of adsorption, the rate of reaction is given by the relationship [7]:

$$r_0 = k_{deg} \frac{K_{LH} C_0}{1 + K_{LH} C_0} \quad (1)$$

where $K_{LH}$ is the adsorption constant, $C_0$ the tylosin concentration in the aqueous phase and $k_{deg}$ a kinetic constant, which depends on various physico-chemical parameters including the irradiation conditions.

Relationship (1) can be rewritten in a linear form as:

$$\frac{C_0}{r_0} = \left(\frac{1}{k_{deg}}\right)\left(\frac{1}{K_{LH}} + C_0\right) \qquad (2)$$

The data are consistent with this linear relationship as shown in Figure 6. From the values of $K_{LH}$ (0.0045 L/mg) and $k_{deg}$ (75.19 mg/L/h), a theoretical curve has been plotted in Figure 5, which agrees with the experimental data. As a consequence, the order with respect to tylosin is near to 1 at low concentration (that is $C_0 < 150$ mg/L), when eq. (1) reduces to eq. (3):

$$r_0 = k_{deg} K_{LH} C_0 \qquad (3)$$

whereas the order tends to a value of 0 at high concentration (beyond the present experimental range) where the rate would reach the limiting value $k_{deg}$.

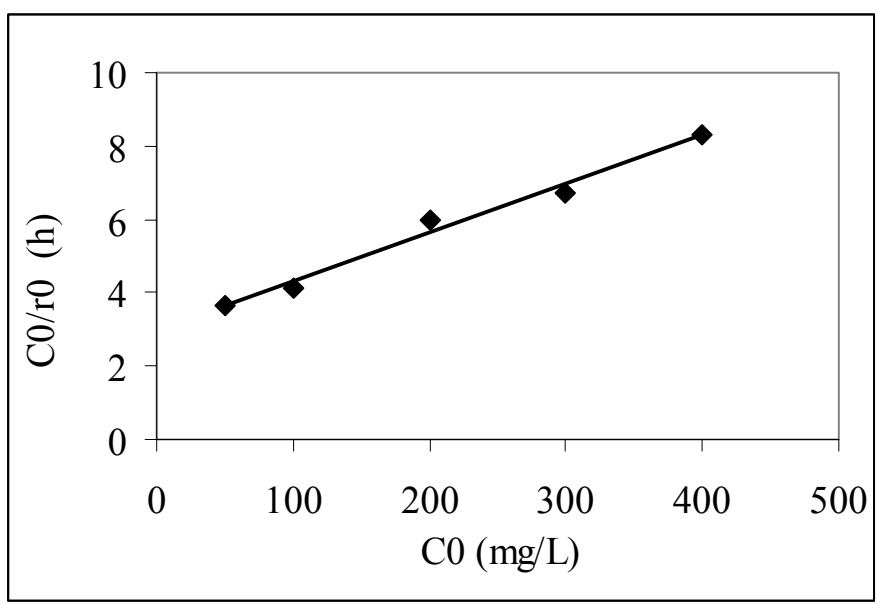

Figure 6 Determination of $k_{deg}$ and $K_{LH}$ constants

If the intermediates do not compete severely with tylosin for adsorption and therefore for reaction with the reactive transient, the kinetics observed at initial time should hold during the reaction time.

## 4   Conclusions

Degradation of tylosin has been successfully conducted by photocatalytic process using immobilized $TiO_2$. The technique is promising as immobilization of the photocatalyst avoids the problem of recovery of the nanoparticles. The effect of tylosin concentration on the photocatalytic oxidation activity was studied, and the kinetics obeys satisfactorily the Langmuir-Hinshelwood model, from which an adsorption and apparent kinetic constants have been determined.